\documentclass[amsmath,amssymb,aps,prb,twocolumn,floatfix,10pt,superscriptaddress]{revtex4-2}

\usepackage{mathtools}
\usepackage[caption = false]{subfig}
\usepackage{graphicx}
\usepackage{standalone}
\usepackage{dcolumn}
\usepackage{bm}
\usepackage{xcolor}
\usepackage{physics}
\usepackage{enumitem}
\usepackage{calligra}
\usepackage[colorlinks = true,linkcolor = red,citecolor = magenta]{hyperref}
\usepackage{natbib}
\usepackage{orcidlink}
\usepackage[normalem]{ulem}
\usepackage{pdfpages}

\makeatletter
\AtBeginDocument{\let\LS@rot\@undefined}
\makeatother

\DeclareGraphicsExtensions{.pdf,.eps,.png,.jpg,.mps}

\newcommand{\pcsadd}{Center for Theoretical Physics of Complex Systems, Institute for Basic Science (IBS), Daejeon, Korea, 34126}
\newcommand{\ustadd}{Basic Science Program, Korea University of Science and Technology (UST), Daejeon 34113, Republic of Korea}
\newcommand{\nziasadd}{Centre for Theoretical Chemistry and Physics, The New Zealand Institute for Advanced
Study (NZIAS), Massey University Albany, Auckland 0745,
New Zealand}

\makeatletter
\renewcommand*{\fnum@figure}{{\normalfont\bfseries \figurename~\thefigure}}
\renewcommand*{\@caption@fignum@sep}{\textbf{:}}
\makeatother

\newcommand{\bea}{\begin{eqnarray}}
\newcommand{\eea}{\end{eqnarray}}

\newcommand{\appropto}{\mathrel{\vcenter{
  \offinterlineskip\halign{\hfil$##$\cr
    \propto\cr\noalign{\kern2pt}\sim\cr\noalign{\kern-2pt}}}}}

\newlength\mylen
\newlist{mycases}{enumerate}{1}
\setlist[mycases,1]{label=\textbf{Case~\arabic*.}, 
  labelwidth=\dimexpr-\mylen-\labelsep\relax,leftmargin=0pt,align=right}

\begin{document}

\title{
Spectral Topology and Delocalization in Disordered Hatano-Nelson Chains
}

\author{Supriyo Ghosh }
    \email{supriyoghosh711@gmail.com}
    \affiliation{Department of Physics, Siksha-Bhavana, Visva-Bharati, Santiniketan 731235, India}
    \affiliation{\pcsadd}



\author{Sergej Flach\,\orcidlink{}}
    \email{sflach@ibs.re.kr}
    \affiliation{\pcsadd}
    \affiliation{\ustadd}
    \affiliation{\nziasadd}

\date{\today}


\begin{abstract}

The unidirectional Hatano-Nelson chain serves as the fundamental non-Hermitian building block of the Su-Schrieffer-Heeger (SSH) model. We investigate its Anderson localization properties under diagonal binary disorder. For weak disorder, the complex eigenvalue spectrum forms a single closed loop, which bifurcates into two distinct loops at a critical disorder threshold. Correspondingly, the spectral winding number $\nu$ undergoes a transition from $\nu=1$ in the weak-disorder regime, through $\nu=1/2$ at the critical point, to $\nu=0$ in the strong-disorder limit. We show that the eigenstates are subexponentially localized, with a localization length that varies analytically as a function of the momentum-like quantum number q. Notably, at weak and critical disorder, the spectrum hosts two completely delocalized states with diverging localization lengths. This divergence is directly correlated with the non-trivial spectral winding number. These findings remain robust under various boundary conditions, with the exception of strictly open boundaries. 
\end{abstract}

\maketitle

\section{Introduction}

Anderson localization (AL)—the confinement of electronic wavefunctions within a limited region of a lattice due to stochastic disorder—remains a cornerstone of condensed matter physics \cite{Anderson1958PR, Anderson1979PRL}. Since its discovery, AL has been extensively explored across diverse physical platforms, including photonic crystal waveguides \cite{Sapienza2010Sc}, light dynamics \cite{Tal2007Nature, Lahini2008PRL}, ultrasound propagation \cite{Weaver1990WM, Hefei2008NP}, and microwave systems \cite{Dalichaouch1991Nature}. Recently, the study of Anderson localization in non-Hermitian systems has emerged as a vibrant frontier \cite{HN1996PRL, HN2016PRE, Huang2020PRB, Jiang2019PRB}. Notably, random non-Hermitian disorder has been shown to induce AL transitions in three-dimensional systems \cite{Huang2020PRB}, while studies in PT-symmetric systems have revealed phase transitions from extended to localized states in optical waveguide arrays with balanced loss and gain (BLG) \cite{Yuce2014PLA, Zeng2017PRA}.

A particularly compelling recent development is the interplay between AL and topological phase transitions \cite{Liu2022PLA, Zuo2022PRA, Altland2015PRA, Quinn2015PRB, Zhang2023PRB, Lin2022NC}. For instance, random binary disorder in the hopping amplitudes of a Su-Schrieffer-Heeger (SSH) model can give rise to topological Anderson insulator phases \cite{Liu2022PLA}. Research in this vein has characterized the system-size scaling of local moments \cite{Quinn2015PRB}, as well as persistent currents and entanglement spectra at topological Anderson transitions \cite{Zhang2023PRB}.

The introduction of non-Hermiticity adds a transformative dimension to these studies. Non-Hermitian topological Anderson insulators have been realized experimentally \cite{Lin2022NC}, and it has been demonstrated that skin states and Anderson localized states can be differentiated by distinct spectral winding numbers \cite{Fortin2025PRB}.

In this article, we investigate the Anderson localization properties and topological phase transitions in a non-Hermitian unidirectional chain subject to diagonal binary disorder. We demonstrate that the localization length diverges under specific conditions dictated by the system's topological phase. Specifically, we find at least two completely extended states within the topologically non-trivial regime. Conversely, in the trivial regime, all eigenstates are localized, and the divergence of the localization length is suppressed.

The paper is organized as follows: In Sec. II.A, we introduce the unidirectional model, followed by a discussion of its symmetries and winding numbers in Sec. II.B. We derive an analytical expression for the localization length in Sec. II.C and analyze participation ratios in Sec. II.D. The impact of generalized boundary conditions (GBC) is examined in Sec. III, and the effects of off-diagonal binary disorder are addressed in Sec. IV. Finally, we summarize our findings in Sec. VI.

\section{Model and results}

We consider a lattice model consisting of \(N\) sites described by the non-Hermitian Hamiltonian:
\begin{eqnarray}
H = \sum_{n=1}^{N} \left( -t_2 \ a_{n}^{\dagger}a_{n+1} + t_1 \ a_{n}^{\dagger}a_{n} \right),
\label{ham1}
\end{eqnarray}
\noindent where \(a_{n}\) (\(a_{n}^{\dagger}\)) is the annihilation (creation) operator at site \(n\). Here, \(t_1\) denotes the onsite potential, while \(t_2\) represents the unidirectional hopping amplitude from site \(n+1\) to site \(n\). Note that the reciprocal hopping amplitude (from site \(n\) to site \(n+1\)) is zero, which serves as the primary source of non-Hermiticity in this lattice model. We assume periodic boundary conditions (PBC) such that \(a_{N+1} \equiv a_{1}\). This Hamiltonian~(\ref{ham1}) represents a unidirectional Hatano-Nelson~\cite{HN1996PRL} chain (uHN) and constitutes the irreducible block of the Su-Schrieffer-Heeger (SSH)~\cite{SSH1979PRL} model. Indeed, the SSH Hamiltonian can be expressed as: 
$H_{SSH} = \frac{1}{2}\left[\left(\sigma_1 + i\sigma_2\right) \otimes H^{\dagger} + \left(\sigma_1 -i\sigma_2\right)\otimes H\right]$.

We define the general eigenstate of the uHN Hamiltonian (\ref{ham1}) as:
\begin{eqnarray}
    \vert \Psi \rangle = \sum_{j = 1}^{N} \psi_{j} \ a_{j}^{\dagger} \ \vert 0 \rangle .
    \label{gen_eig_state}
\end{eqnarray}
The Schrödinger equation \(H\vert \Psi \rangle = E\vert \Psi \rangle\) yields the following set of coupled equations:
\begin{eqnarray}
    E\psi_{n} & = & t_1 \ \psi_{n} - t_2 \ \psi_{n+1} \ ;\hspace{2mm} n = 1,2 \dots (N-1), \nonumber \\
    E\psi_{N} & = & t_1 \ \psi_{N} - t_2 \ \psi_{1}.
    \label{set_schr}
\end{eqnarray}
By employing the ansatz \(\psi_{n} = A \ e^{ikn}\), we obtain the energy dispersion \(E = t_1 - t_2 e^{ik}\). The boundary conditions restrict the allowed values of the wavevector to \(k = \frac{2\pi s}{N}\), where \(s = 0,1,\dots (N-1)\). In the complex energy plane, the eigenvalue spectrum of the ordered uHN model forms a circle centered at $t_1$ on the real axis with radius $t_2$. The corresponding eigenvectors are extended plane waves, where the wavevector $k$ corresponds to the angular position of a point on the spectral circle.

\subsection{Binary Onsite Disorder}

To investigate Anderson localization within the non-Hermitian unidirectional Hatano-Nelson (uHN) lattice model, we introduce binary disorder into the onsite potential:
\begin{eqnarray}
H = \sum_{n=1}^{N} \left(t_{1n} \ a_{n}^{\dagger}a_{n} - t_2 \ a_{n}^{\dagger}a_{n+1} \right), \quad t_{1n} \in \{+h, -h\}.
\label{main_ham}
\end{eqnarray}
The onsite potential $t_{1n}$ assumes uncorrelated values of $+h$ or $-h$ randomly with equal probability. Considering the thermodynamic limit with a large even $N$, we assume that half of the sites possess a potential $+h$ and the other half $-h$. In this regime, the eigenvalues can be derived analytically.

We obtain a system of equations similar to Eq.~(\ref{set_schr}), but with the distinction that $t_1$ fluctuates between $\pm h$. This leads to the ratios $\frac{\psi_{n+1}}{\psi_{n}} = -\frac{E-h}{t_2}$ or $\frac{\psi_{n+1}}{\psi_{n}} = -\frac{E+h}{t_2}$. Consequently, the boundary condition for the periodic chain requires:
\begin{eqnarray}
    \prod_{n=1}^{N} \frac{\psi_n}{\psi_{n+1}} = \frac{t_2^N}{(E^2 - h^2)^{N/2}} \implies 1 = \frac{t_2^N}{(E^2 - h^2)^{N/2}}. \label{TM1}
\end{eqnarray}
The resulting eigenvalues are given by:
\begin{eqnarray}
    E_{\pm} = \pm \sqrt{h^2 + t_2^2 \ e^{iq}},
\end{eqnarray}
where $q = \frac{4\pi s}{N}$ and $s = 0, 1, \dots, \frac{N}{2} - 1$. In polar form, the eigenvalues can be expressed as:
\begin{eqnarray}
    E_{\pm} = \pm \sqrt{R} \ e^{i\phi/2},
\end{eqnarray}
where
\begin{eqnarray}
    R = \sqrt{h^4 + t_2^4 + 2h^2t_2^2\cos(q)}, \quad \phi = \arctan\left(\frac{t_2^2 \sin(q)}{h^2 + t_2^2 \cos(q)}\right). \nonumber
\end{eqnarray}

\begin{figure}[htbp]
\centering
    \subfloat[]{\includegraphics[width = 0.333\linewidth]{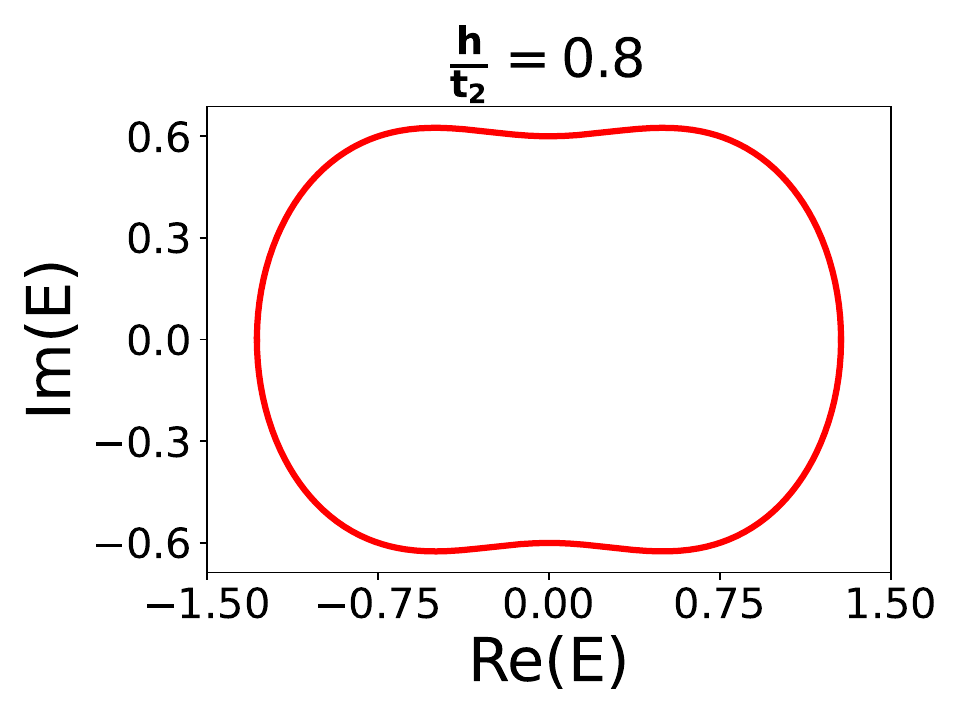}}
    \subfloat[]{\includegraphics[width = 0.333\linewidth]{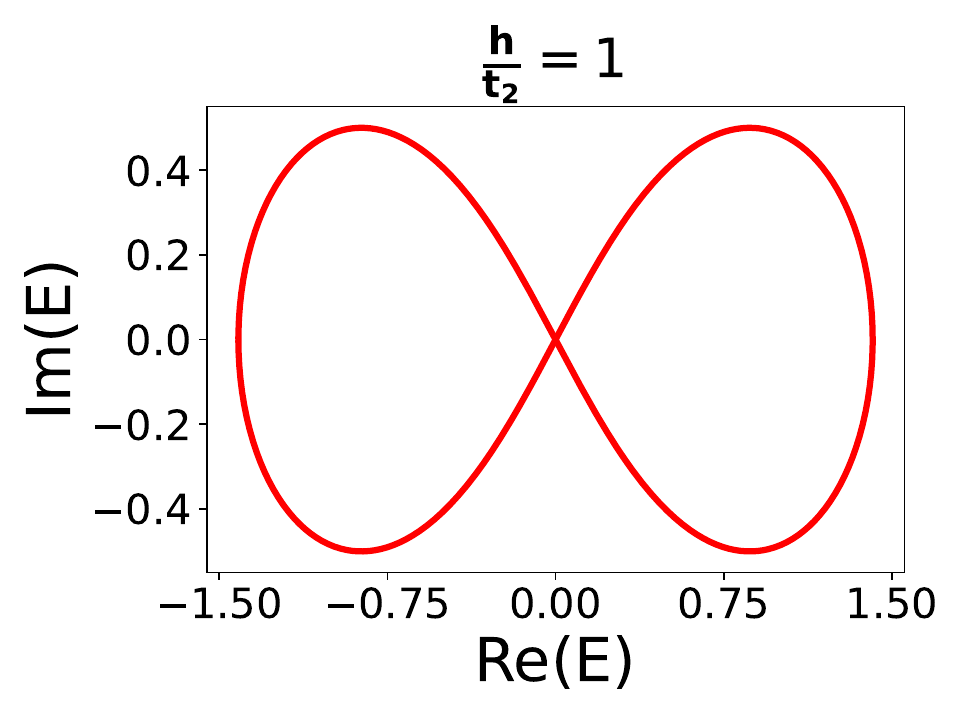}}
    \subfloat[]{\includegraphics[width = 0.333\linewidth]{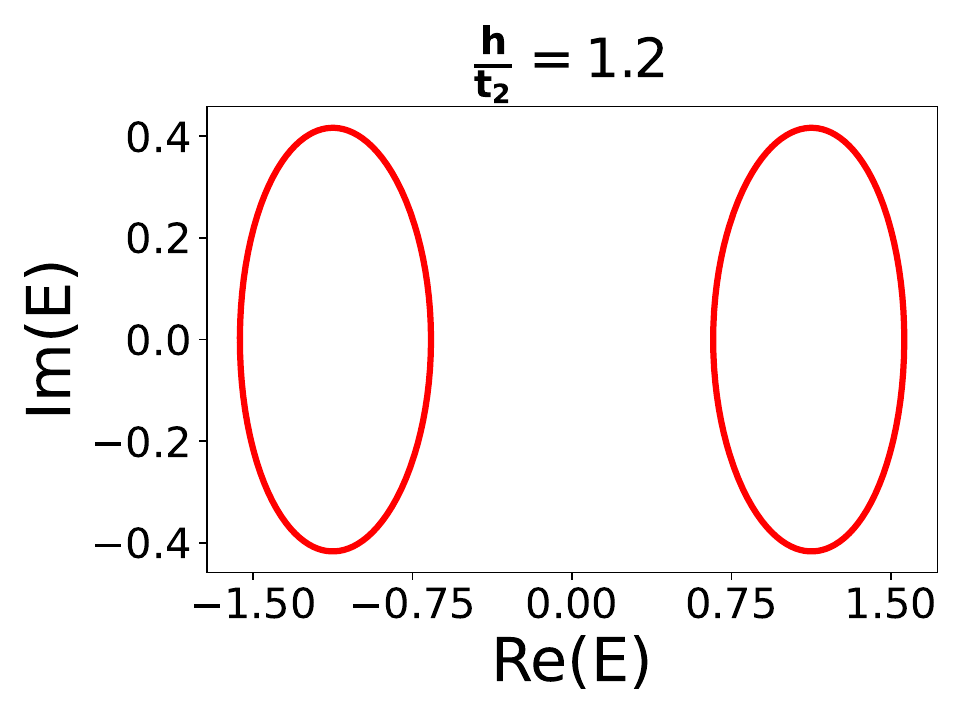}}
    \caption{Complex energy spectra of the binary disordered uHN chain. Parameters: $t_2 = 1, N \to \infty$. (a) Weak disorder ($h = 0.8 < t_2$); (b) Critical point ($h = 1.0 = t_2$); (c) Strong disorder ($h = 1.2 > t_2$).}
    \label{spectrum_pbc}
\end{figure}

The complex spectra are illustrated in Fig.~\ref{spectrum_pbc}. For weak disorder ($h < t_2$), the eigenvalue spectrum forms a single closed loop in the complex plane, which bifurcates into two distinct loops at the critical disorder strength $h = t_2$. The expression for the spectrum and the bifurcation of the spectral loop were reported in Ref. \cite{Feinberg1999PRE}, while the spectral topology and localization length were not. In the following sections, we investigate the spectral topology and the Anderson transition. As discussed in the following section, the spectral winding number undergoes a transition from $1$ to $0$ via the value $1/2$ at the critical point as the system moves from the weak to the strong disorder regime \footnote{These spectral results were first introduced to us by N. Poovuttikul et al (unpublished), see also detailed acknowlegement.}.

For $t_2 > h$, the eigenvalues corresponding to $q = \pi$ are purely imaginary. For $h \leq t_2$ and $q = \pi$, the eigenvalue is $E_{\pm} = \pm i \sqrt{t_2^2 - h^2}$. In this case, the Schrödinger equation becomes:
\begin{eqnarray}
\left(\pm i \sqrt{t_2^2 - h^2} - t_{1n} \right) \psi_{n} = - t_2 \psi_{n+1} .
\end{eqnarray}
This implies that $|\psi_{n+1}| = |\psi_{n}|$, a condition of uniform amplitude that also holds at the critical point $t_2 = h$. Normalization ensures that $|\psi_{n}| = 1/\sqrt{N}$. The phase difference between $\psi_{n+1}$ and $\psi_{n}$ for $h \leq t_2$ at $q = \pi$ is:
\begin{eqnarray}
\Delta_{n} & = & \mp \arctan \left(\frac{\sqrt{t_2^2 - h^2}}{h}\right) \quad \text{for } t_{1n} = +h, \nonumber \\
           & = & \pm \left[\arctan \left(\frac{\sqrt{t_2^2 - h^2}}{h}\right) \pm \pi \right] \quad \text{for } t_{1n} = -h, \nonumber 
\end{eqnarray}
where the $\pm$ sign corresponds to $E = E_{\pm}$. These analytical results demonstrate that for $h \leq t_2$, the $q = \pi$ eigenstates are completely delocalized with constant absolute values and phase shifts directly imprinted by the specific disorder realization. Conversely, numerical diagonalization reveals that all other eigenstates are localized.

\begin{figure}[htbp]
\centering
    \subfloat[]{\includegraphics[width = 0.49\linewidth]{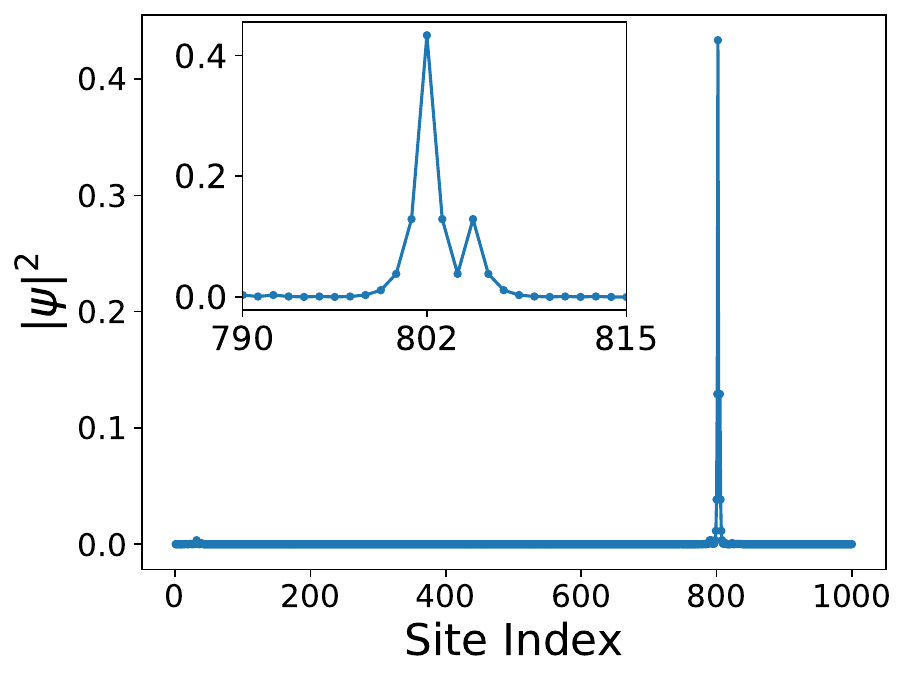}}
    \subfloat[]{\includegraphics[width = 0.49\linewidth]{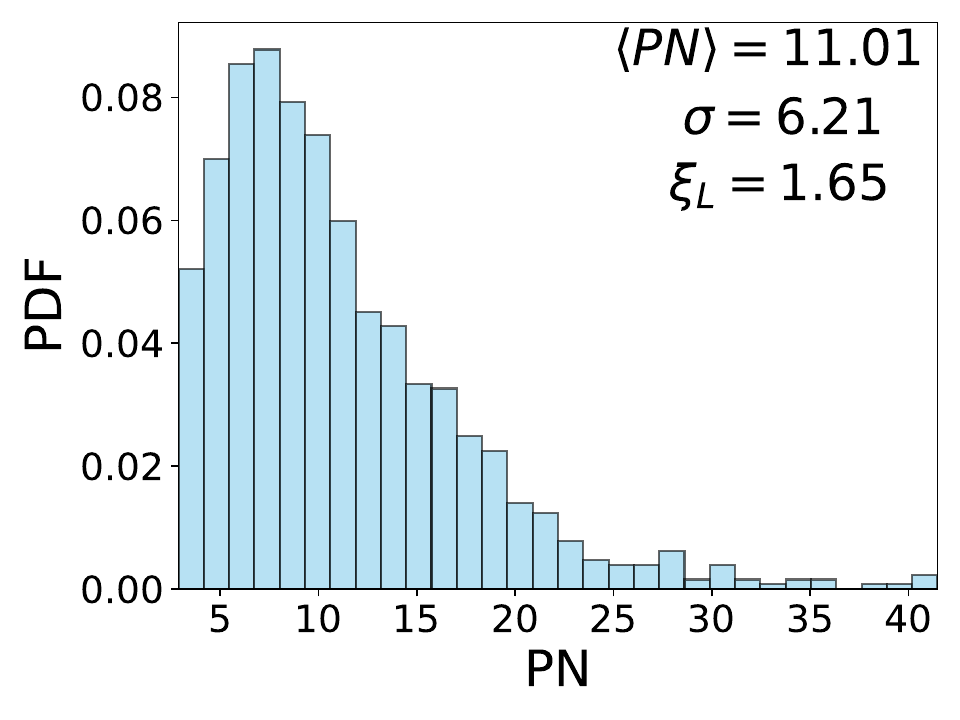}}
    \caption{(a) Spatial profile of $|\psi|^2$ for a localized eigenstate. (b) Probability density function (PDF) of the participation number for 1000 random disorder realizations. Here,`PN’ stands for participation number as shown in Eq.~(\ref{PN_Eqn}). Parameters: $t_2 = 1, h = 0.8, N = 1000, q = \pi/2$.}
    \label{abs_wf}
\end{figure}

The squared absolute value of a typical localized eigenstate is shown in Fig.~\ref{abs_wf}(a), illustrating its spatial confinement in real space.

\subsection{Symmetries and Winding Number}

The Hamiltonian exhibits chiral symmetry, time-reversal symmetry, and particle-hole symmetry. We define a unitary operator $S$ as $S = I_{N/2} \otimes \sigma_{z}$. Under the transformation $SHS^{-1} = 2t_1 - H$, and given that the onsite potential is binary ($t_{1n} \in \{h, -h\}$), the Hamiltonian satisfies chiral symmetry: $SHS^{-1} = -H$. This ensures that if $E$ is an eigenvalue, $-E$ must also be an eigenvalue, consistent with the relation $E_{+} = -E_{-}$.

Furthermore, the Hamiltonian possesses time-reversal symmetry with respect to the operator $\mathcal{T} = I_{N}\mathcal{K}$, where $\mathcal{K}$ is the complex conjugation operator. The invariance $\mathcal{T}H\mathcal{T}^{-1} = H$ indicates that if $E$ is an eigenvalue, its complex conjugate $E^{*}$ is also an eigenvalue. Similarly, particle-hole symmetry exists with respect to the operator $\mathcal{C} = S\mathcal{K}$. The transformation $\mathcal{C}H\mathcal{C}^{-1} = -H$ implies that if $E$ is an eigenvalue, then $-E^{*}$ is also an eigenvalue.

In non-Hermitian systems, the energy spectrum is generally complex, and conventional band gaps are not well defined. Consequently, the topology of such systems is characterized by the winding of the complex energy spectrum in the complex plane. For a translationally invariant non-Hermitian system, the spectral winding number is defined as
\begin{eqnarray}
    W = \frac{1}{2\pi i} \oint dk \frac{d}{dk} \ln \det\left[H(k) - E_0\right] 
\end{eqnarray}
where $k \in \left[-\pi, \pi\right]$ is the wave-vector and $E_0$ is a reference energy in the complex plane that lies outside the spectrum. 

However, in the presence of disorder, translational symmetry is broken, and the wave vector $k$ is no longer a good quantum number. In such cases, the topology of the non-Hermitian disordered system is characterized by the winding of the complex spectrum under the variation of an external flux, which effectively replaces the role of momentum in defining the spectral winding number. The eigenvalue winding number \cite{Gong2018PRX} for a disordered system is defined as:
\begin{equation}
\nu(E_0) = \frac{1}{2i\pi} \int_{0}^{2\pi} \frac{d}{d\Phi} \ln \det \left[ H(\Phi) - E_{0} \right] d\Phi,
\label{winding_def}
\end{equation}
where $H(\Phi)$ represents the Hamiltonian subject to an external Aharonov-Bohm flux $\Phi$, expressed as:
\begin{equation}
H(\Phi) = \sum_{n=1}^{N} \left( t_{1n} a_{n}^{\dagger}a_{n} - t_2 e^{i\Phi/N} a_{n}^{\dagger}a_{n+1} \right), \quad t_{1n} \in \{+h, -h\}.
\end{equation}

In our model, the expression for the winding number around $E_0 = 0$ simplifies to:
\begin{equation}
\nu(0) = \frac{1}{2 i\pi} \oint_{\mathcal{C}} \frac{dz}{z - (-1)^{N/2} (h/t_2)^{N}},
\end{equation}
where $\mathcal{C}$ denotes the unit circle contour $z = e^{i\Phi}$ in the complex plane.

The presence of a pole inside, on, or outside the contour depends on the ratio of $h$ to $t_2$. Specifically, the winding number at $E_0 = 0$ is found to be:
\begin{eqnarray}
    \nu(0) = 
    \begin{cases} 
      0 & (h > t_2) \\
      1/2 & (h = t_2) \\
      1 & (h < t_2) 
    \end{cases}
\end{eqnarray}
Thus, the spectral winding number transitions from a topologically non-trivial regime ($\nu(0) = 1$) to a trivial regime ($\nu(0) = 0$) via a critical point ($\nu(0) = 1/2$) as the disorder strength $h$ increases relative to the hopping amplitude $t_2$.

\subsection{Localization Length}

With the use of (\ref{TM1}) we can obtain the propagation of the log of the absolute value of the wave function of an eigenstate at given eigenvalue $E$ as $\ln |\psi_{n+1}| = \ln |\psi_n| + \ln |\frac{E-t_{1n}}{t_2}|$. This is random walk with zero bias (due to the imposed boundary conditions) which leads to subexponential localization $\psi_n \sim {\rm e}^{-\sqrt{\gamma n}}$ as shown by Silvestrov \cite{Silvestrov1998PRB}. The characteristic localization length scale $\xi_L = 1/\gamma$ is in general of finite value. We also note here for clarity that non-Hermitian matrices have different left and right eigenvectors, with us here focusing on right eigenvectors only.

The local inverse localization length can be defined as $\gamma_n = \pm \ln |\psi_{n+1}/\psi_n|$, where the sign corresponds to the decay on the left or right side of the peak. Focusing on the absolute decay rate, we define the average inverse localization length in the thermodynamic limit as:
\begin{eqnarray}
    \gamma & = & \lim_{L \to \infty} \frac{1}{L}\sum_{n=1}^{L} \gamma_{n} \nonumber \\ 
           & = & \lim_{L \to \infty} \frac{1}{L}\sum_{n=1}^{L} \left| \ln \left| \frac{\psi_{n+1}}{\psi_{n}} \right| \right|.
\end{eqnarray}
As established in the previous section, the ratio $|\psi_{n+1}/\psi_{n}|$ assumes values of $|(E+h)/t_2|$ or $|(E-h)/t_2|$ with equal probability. In the limit $L \to \infty$, the average inverse localization length becomes:
\begin{equation}
    \gamma = \frac{1}{2} \left( \left| \ln \left| \frac{E+h}{t_2} \right| \right| + \left| \ln \left| \frac{E-h}{t_2} \right| \right| \right).
\end{equation}
Given the property $|(E+h)/t_2| \cdot |(E-h)/t_2| = 1$, both logarithmic terms contribute equally. Thus, the final expression for the inverse localization length is:
\begin{eqnarray}
    \gamma & = & \left| \ln \left| \frac{E+h}{t_2} \right| \right| \nonumber \\
           & = & \frac{1}{2} \ln \left( \frac{R + h^2 + 2\sqrt{R}h \cos(\phi/2)}{t_2^2} \right).
\end{eqnarray}
The localization length is subsequently retrieved via the relation $\xi_{L} = 1/\gamma$. 

The eigenvalues corresponding to $E_{+}$ and $q \in [0, \pi]$ constitute the irreducible part of the spectrum. At $q = 0$ (where $\phi = 0$), the inverse localization length simplifies to $\gamma = \ln[(h + \sqrt{t_2^2 + h^2})/t_2]$. At the Brillouin zone boundary $q = \pi$, the behavior depends on the disorder strength: for $h \leq t_2$, $\gamma$ vanishes, implying that $\xi_{L} \to \infty$. This indicates that the state corresponding to $q = \pi$ is a completely extended state. Conversely, for $h > t_2$, $\gamma = \frac{1}{2} \ln[(2h^2 - t_2^2 + 2h\sqrt{h^2 - t_2^2})/t_2^2]$, which remains finite.

Consequently, the existence of extended states is fundamentally tied to the spectral winding number. In the parameter regime of non-trivial topology, at least two states are completely extended and exhibit a diverging localization length. To investigate the scaling behavior near the delocalization point, we Taylor expand $\gamma$ near $q = \pi$ as a function of $g = h/t_2$:
\begin{widetext}
\begin{eqnarray}
    \gamma & \approx & \frac{1}{2} \left| -\frac{g}{\sqrt{1-g^2}} (q - \pi) + \frac{g^3}{6(1-g^2)^{3/2}} (q - \pi)^3 + \dots \right| \quad (g < 1) \nonumber \\
           & \approx & \frac{1}{2} \left| \sqrt{2} |q-\pi|^{1/2} - \frac{1}{2} |q - \pi| + \frac{1}{2\sqrt{2}} |q - \pi|^{3/2} \dots \right| \quad (g = 1) \nonumber \\
           & \approx & \frac{1}{2} \left| \ln(2g^2 - 1 + 2g\sqrt{g^2 - 1}) + \frac{g (q - \pi)^{2}}{4(g^2 - 1)^{3/2}} + \dots \right| \quad (g > 1)
\end{eqnarray}
\end{widetext}
where $g = \frac{h}{t_2}$.

\subsection{Participation Number}

To quantify the degree of localization in our system, we calculate the participation number ($PN$) for each eigenstate. For the $n$-th state, the participation number is defined as:
\begin{equation}
    PN_{n} = \frac{\left( \sum_{j=1}^{N} |\psi_{j,n}|^2 \right)^2}{\sum_{j=1}^{N} |\psi_{j,n}|^4},
    \label{PN_Eqn}
\end{equation}
where $j$ and $n$ denote the lattice site index and the eigenvalue index, respectively. For a completely delocalized state distributed uniformly across the lattice, the wavefunction amplitude is $|\psi_{j}| \approx 1/\sqrt{N}$, leading to a participation number $PN \approx N$. Conversely, for a state localized at a single site, $PN \approx 1$.

We numerically evaluate the participation number for the eigenstates corresponding to $E = E_{+}$ across the irreducible Brillouin zone $q \in [0, \pi]$. The numerical procedure involves constructing the Hamiltonian matrix with a diagonal binary disorder, ensuring an equal number of $+h$ and $-h$ onsite potentials to satisfy the half-filling condition. After numerical diagonalization, we map the numerical eigenstates to their corresponding wavevectors $q$ by comparing the numerical eigenspectrum with our analytical dispersion relation. For each $q$, the $PN$ is calculated using Eq.~(\ref{PN_Eqn}) and averaged over 100 independent disorder realizations to ensure statistical convergence.

\begin{figure}[htbp]
\centering
    \subfloat[]{\includegraphics[width = 0.333\linewidth]{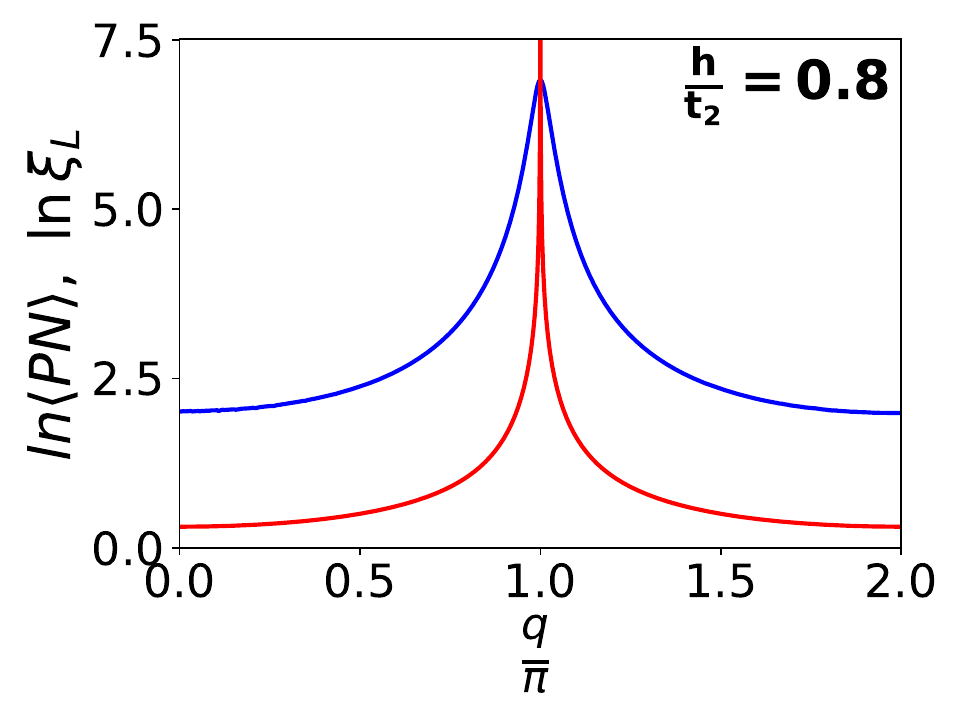}}
    \subfloat[]{\includegraphics[width = 0.333\linewidth]{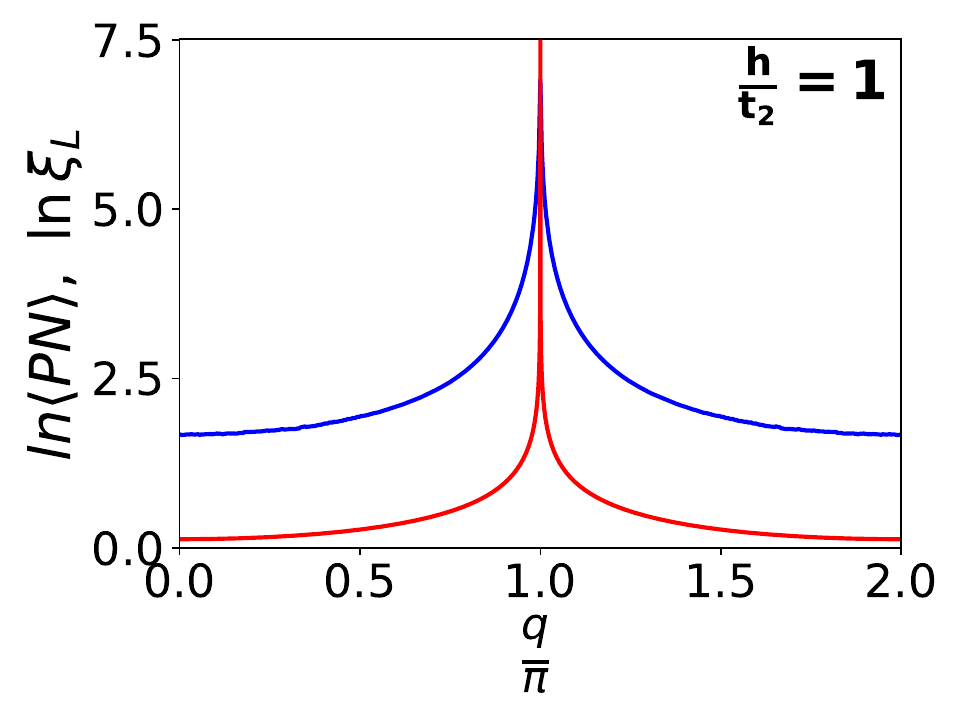}}
    \subfloat[]{\includegraphics[width = 0.333\linewidth]{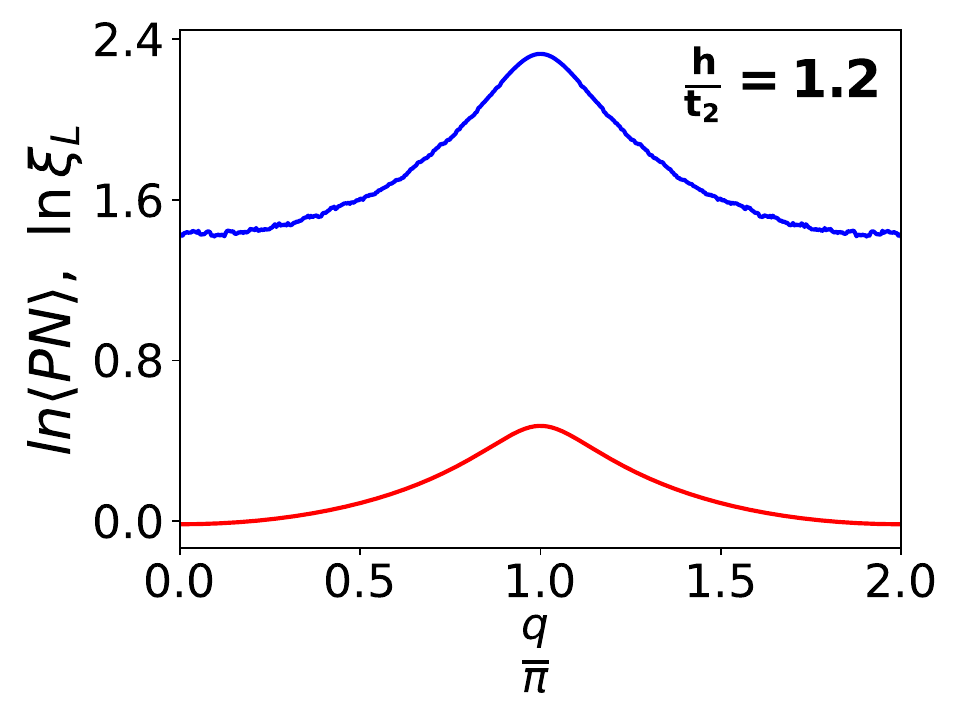}}
    \caption{Participation number (blue solid lines) and analytical localization length (red solid lines) as a function of $q$. The participation number is averaged over 100 random disorder configurations. Parameters: $t_2 = 1$, $N = 1000$; (a) weak disorder ($h = 0.8$); (b) critical disorder ($h = 1.0$); (c) strong disorder ($h = 1.2$).}
    \label{PN_pbc}
\end{figure}

The numerical results for the participation number and the analytical localization length are presented in Fig.~\ref{PN_pbc} for the weak, critical, and strong disorder regimes. The plots clearly show that at the state $q = \pi$, the participation number approaches the system size $N$ for both the weak disorder ($\nu(0) = 1$) and critical ($h = t_2$) regimes. These numerical results provide strong confirmation of our analytical findings, specifically the existence of topologically protected delocalized states at $q = \pi$ within the non-trivial winding phase.

The correlation between the topological transition and existence of delocalized states could be explained as follows : 
The introduction of an external AB flux through a one-dimensional ring generates an electromotive force in the system. States with delocalized wavefunctions are sensitive to the boundary phase $\Phi$, and their energies evolve continuously as the flux is varied. As a result, the complex eigenvalues trace trajectories in the complex energy plane as $\Phi$ is tuned, giving rise to a non-zero spectral winding number.

In contrast, localized states are insensitive to the boundary phase, since their wavefunctions decay before reaching the boundary. Consequently, their eigenvalues remain essentially unchanged under variations of $\Phi$. Therefore, if all eigenstates are localized, the spectrum does not wind in the complex plane and the spectral winding number vanishes.

\section{Generalized Boundary Conditions}

To examine the influence of boundary conditions on the localization and topological properties of the system, we consider the Hamiltonian of the unidirectional lattice model with generalized boundary conditions (GBC):
\begin{eqnarray}
    H = -t_2 \sum_{n=1}^{N-1} a^{\dagger}_{n}a_{n+1} + \sum_{n=1}^{N} t_{1n} a^{\dagger}_{n}a_{n} - \alpha t_2 \ a_{N}^{\dagger} a_{1},
\end{eqnarray}
where $0 \leq \alpha \leq 1$ and the onsite potential remains binary, $t_{1n} \in \{+h, -h\}$. The parameter $\alpha$ tunes the boundary configuration: $\alpha = 0$ corresponds to open boundary conditions (OBC), while $\alpha = 1$ retrieves the periodic boundary conditions (PBC).

Using methods analogous to those employed for PBC, we analytically derive the eigenspectra for the GBC. The characteristic polynomial equation for this system is given by:
\begin{equation}
    (E^2 - h^2)^{N/2} = \alpha \ t_2^{N}.
\end{equation}
Solving for $E$, the expression for the eigenvalues is:
\begin{equation}
    E_{\pm} = \pm \sqrt{h^2 + t_2^2 \alpha^{2/N} e^{iq}},
\end{equation}
where $q = \frac{4\pi s}{N}$ for $s = 0, 1, \dots, (\frac{N}{2}-1)$.

In the limit of strictly open boundaries ($\alpha = 0$), the energy bands become completely flat at $E_{\pm} = \pm h$. However, for any non-zero value of $\alpha$, the eigenvalues converge to the PBC spectrum as $N \to \infty$ due to the scaling of the term $\alpha^{2/N}$. Since the localization length $\xi_L$ is defined in the thermodynamic limit ($N \to \infty$), its value remains identical to that obtained under PBC. Consequently, the emergence of Anderson localization and the associated topological phase transition are robust features of the system, remaining invariant under changes in the boundary conditions, with the exception of the singular OBC case.

\section{Binary Disorder in the Hoppings}

To compare the effects of different types of disorder, we investigate binary disorder within the hopping amplitudes. In this configuration, the Hamiltonian is expressed as:
\begin{equation}
H = \sum_{n=1}^{N} \left(t_{1} \ a_{n}^{\dagger}a_{n} - t_{2n} \ a_{n}^{\dagger}a_{n+1} \right), \quad t_{2n} \in \{+h, -h\}.
\label{dis_hop_ham}
\end{equation}
By solving the characteristic equation for this disordered system, we obtain the following analytical expressions for the eigenvalues:
\begin{eqnarray}
    E & = & t_1 + h e^{i \frac{2\pi s}{N}} \quad \text{for } N/2 \text{ even}, \nonumber \\
      & = & t_1 + h e^{i \frac{\pi (2s+1)}{N}} \quad \text{for } N/2 \text{ odd},
\end{eqnarray}
where $s = 0, 1, \dots, N-1$. The corresponding Schrödinger equation for each site $n$ is given by:
\begin{equation}
    (E - t_1) \psi_{n} = - t_{2n} \psi_{n+1}.
\end{equation}

From this relation, it immediately follows that $|E - t_1| |\psi_{n}| = |t_{2n}| |\psi_{n+1}|$. Since $|t_{2n}| = h$ for all $n$, and $|E - t_1| = h$ for all eigenvalues in the spectrum, we arrive at the condition:
\begin{equation}
    |\psi_{n+1}| = |\psi_{n}|.
\end{equation}
This condition holds for all eigenvalues, implying that every eigenstate in the system remains completely extended despite the presence of binary disorder in the hopping terms.

In the standard Hermitian Su-Schrieffer-Heeger (SSH) model, interchanging the intracell and intercell hopping amplitudes yields a symmetric set of eigenvalues. However, in this irreducible non-Hermitian block, binary disorder in the onsite potential and the hopping amplitude produce fundamentally different physical outcomes: while onsite disorder drives a topological transition and localizes a majority of the spectrum, hopping disorder is incapable of inducing Anderson localization.

\section{Connection to the SSH Model}

The Hamiltonian described in Eqs.~(\ref{ham1}) and (\ref{main_ham}) serves as the irreducible block of the Su-Schrieffer-Heeger (SSH) model. Mapping the unidirectional Hatano-Nelson block onto the SSH model physically corresponds to introducing binary disorder into the intra-cell hopping amplitudes. The Schrödinger equation for the full SSH system is given by:
\begin{equation}
    \begin{pmatrix} 0 & H^{\dagger} \\ H & 0 \end{pmatrix} \begin{pmatrix} \psi_A \\ \psi_B \end{pmatrix} = E \begin{pmatrix} \psi_{A} \\ \psi_{B} \end{pmatrix},
\end{equation}
where $\psi_{A}$ and $\psi_{B}$ represent the eigenstates associated with the $A$ and $B$ sublattices, respectively. This system can be decoupled into two independent eigenvalue problems:
\begin{equation}
    HH^{\dagger} \psi_{B} = E^2 \psi_{B}, \quad H^{\dagger}H \psi_{A} = E^2 \psi_A.
\end{equation}

Thus, the eigenstates and eigenvalues of the standard SSH model are related to the irreducible block through the singular value decomposition (SVD), $H = \tilde{U}\Sigma \tilde{V}^{\dagger}$. The columns of the left-singular matrix $\tilde{U}$ provide $\psi_{B}$, the columns of the right-singular matrix $\tilde{V}$ provide $\psi_{A}$, and the diagonal elements of $\Sigma$ yield the singular values, which correspond to the energy $E$ of $H_{SSH}$. It is important to note that the eigenstates of $H$ itself are not directly identical to the eigenstates of the SSH model.

A critical question arises: can binary disorder induce localization in the SSH model? To address this, we analyze the operator $HH^{\dagger}$ in the site basis:
\begin{equation}
    HH^{\dagger} = \sum_{n=1}^{N} \left[(t_2^2 + h^2) |n\rangle\langle n| - \mu_{n+1} ht_2 \left(|n\rangle \langle n+1 | + |n+1\rangle \langle n | \right) \right],
\end{equation}
where $\mu_{n} = \text{sgn}(t_{1n}) \in \{+1, -1\}$ and we assume periodic boundary conditions ($N+1 \equiv 1$). The Schrödinger equation $HH^{\dagger}|\Psi\rangle = \tilde{E} |\Psi\rangle$ yields the following recurrence relation:
\begin{equation}
    (\tilde{E} - h^2 - t_2^2) \psi_{n} = -ht_2 (\mu_{n+1} \psi_{n+1} + \mu_{n} \psi_{n-1}).
\end{equation}

In one-dimensional systems, it is well established that this specific type of binary hopping disorder can be gauged away via a unitary transformation. We define $\psi_{n} = \left(\prod_{j=2}^{n} \mu_{j}\right) \psi_{n}'$ for $n \geq 2$, with $\psi_1 = \psi_1'$. Under this transformation, the bulk Schrödinger equation ($n = 2, \dots, N-1$) becomes:
\begin{equation}
    (\tilde{E} - h^2 - t_2^2) \psi_{n}' = -ht_2 (\psi_{n+1}' + \psi_{n-1}').
\end{equation}
The boundary equations are similarly transformed:
\begin{eqnarray}
    (\tilde{E}- h^2 - t_2^2) \psi_{1}'  & = & -ht_2 (\psi_{2}' + P \psi_{N}'), \nonumber \\
    (\tilde{E} - h^2 - t_2^2) \psi_{N}' & = & -ht_2 (\psi_{N-1}' + P\psi_{1}'),
\end{eqnarray}
where $P = \prod_{j=1}^{N} \mu_{j}$ is the parity of the disorder realization. This transformed system describes a tight-binding chain with uniform hoppings.

The resulting boundary conditions are either periodic ($P=1$) or anti-periodic ($P=-1$). Using the plane-wave ansatz $\psi_{n}' = e^{ink}$, the eigenvalues are found to be $\tilde{E} = h^2 + t_2^2 - 2ht_2 \cos k$, where the allowed wavevectors are:
\begin{eqnarray}
    k = 
    \begin{cases} 
      \frac{2s\pi}{N} & \text{for } P = 1 \\
      \frac{(2s+1)\pi}{N} & \text{for } P = -1 
    \end{cases}
\end{eqnarray}
for $s = 0, 1, \dots, N-1$. Since the system with $\pm h$ binary hopping disorder is unitarily equivalent to a uniform tight-binding model, the disorder cannot induce localization in the eigenstates of $HH^{\dagger}$, and consequently, the SSH model remains in an extended phase under this type of binary disorder.

\section{Conclusion and Discussion}

In summary, we have investigated non-Hermitian Anderson localization and the associated topological phase transition in a unidirectional Hatano-Nelson (uHN) chain subject to binary onsite disorder. Our results demonstrate that for weak disorder, the eigenvalue spectrum forms a single closed loop in the complex energy plane. As the disorder strength exceeds a critical threshold, this spectrum bifurcates into two distinct loops. Correspondingly, the spectral winding number $\nu$ undergoes a transition from $\nu = 1$ to $\nu = 0$, passing through the critical value $\nu = 1/2$ at the transition point.

A key finding is the emergence of two purely imaginary eigenvalues when the spectral winding number is non-zero. The corresponding eigenstates are found to be completely delocalized, characterized by a diverging localization length. We have analytically derived the wavefunctions for these extended states and shown that in the non-trivial topological regime, the localization length $\xi_L$ tends to diverge as the wavevector $q \to \pi$. In contrast, when the winding number vanishes, all eigenstates in the spectrum become localized. This establishes a direct correlation between the spectral topology and the divergence of the localization length.

Finally, we examined the system under generalized boundary conditions (GBC). In the case of strictly open boundary conditions (OBC), the spectrum collapses into two flat bands at $E(q) = \pm h$. However, for any other GBC configuration, the eigenspectrum converges to the periodic boundary condition (PBC) result in the thermodynamic limit. This confirms that the observed topological transition and the existence of delocalized states are robust features that remain invariant under variations of the boundary conditions, with the singular exception of the open boundary case.

Future research directions include the impact of the extension to other types of disorder (see e.g. \cite{longhi2021spectral}) on such central results as topology and delocalization, among others.

\begin{acknowledgements}
This work was made possible thanks to crucial initializations and detailed discussions with Dr Napat Poovuttikul and his group. We are very much indebted to them for sharing their research insights with us.
    The authors acknowledge financial support from the Institute for Basic Science (IBS) in the Republic of Korea through Project No. IBS-R024-D1. We thank Yeongjun Kim and Alexei Andreanov for useful discussions.
    \\
    {\sl SF thanks the issue editors for the opportunity to dedicate this piece of work to the memory of Johannes Richter. Johannes was a true friend and scientist, sharing many joyful moments in life and on different continents, spanning times from the 1980s to 2020s and spaces from Germany to Korea to New Zealand. }
\end{acknowledgements}

\bibliography{ref}
\end{document}